\documentclass[twocolumn,aps,pre,floatfix,superscriptaddress]{revtex4-1}
\usepackage{graphicx}
\usepackage{epic,eepic}
\usepackage{graphics}
\usepackage{epsfig}
\usepackage{subfigure}

\usepackage[dvipsnames]{xcolor}

\newcommand{\be}{\begin{equation}}
\newcommand{\ee}{\end{equation}}
\newcommand{\bea}{\begin{eqnarray}}
\newcommand{\eea}{\end{eqnarray}}

\usepackage{amsmath}
\usepackage{amsfonts}
\usepackage{amssymb}
\usepackage{bm}
\usepackage{color}
\usepackage{graphicx}
\usepackage[utf8x]{inputenc}

\begin{document}
%\doublespacing
%\setstretch{2.}
\title{A lattice Boltzmann model for self-diffusiophoretic particles near and at liquid-liquid interfaces}

\author{Lucas S. Palacios}
 \affiliation{Institute for Bioengineering of Catalonia (IBEC), The Barcelona Institute of Science and Technology (BIST), Baldiri i Reixac 10-12, 08028 Barcelona, Spain.}
\author{Andrea Scagliarini}
\affiliation{Istituto per le Applicazioni del Calcolo, CNR - Via dei Taurini 19, 00185 Rome, Italy}
\affiliation{INFN, Sezione di Roma "Tor Vergata", Via della Ricerca Scientifica 1, 00133 Rome, Italy}
\author{Ignacio Pagonabarraga}
 \email{ipagonabarraga@ub.edu}
\affiliation{Departament de F\'{\i}sica de la Materia Condensada, Universitat de Barcelona, Carrer Mart\'{\i} i Franqu\'es 1, 08028 Barcelona, Spain}
\affiliation{Universitat de Barcelona Institute of Complex Systems (UBICS), Universitat de Barcelona, 08028 Barcelona, Spain}
\affiliation{CECAM, Centre Europe\'en de Calcul Atomique et Mol\'eculaire, Ecole Polytechnique F\'ed\'erale de Lausanne (EPFL), Avenue Forel 2, 1015 Lausanne, Switzerland}

\begin{abstract}
We introduce a novel mesoscopic computational model based on a multiphase-multicomponent lattice Boltzmann method for the simulation of self-phoretic particles in the presence of liquid-liquid interfaces. Our model features fully resolved solvent hydrodynamics and, thanks to its versatility, it 
can handle important aspects of the multiphysics of the problem, including particle wettability and differential solubility of the product in the 
two liquid phases.
The method is extensively validated in simple numerical experiments, whose outcome is theoretically predictable, and then applied to the study
of the behaviour of active particles next to and trapped at interfaces. We show that their motion can be variously steered by tuning 
relevant control parameters, such as the phoretic mobilities, the contact angle and the product solubility.
\end{abstract}

\maketitle

\section{Introduction}\label{sec:Intro}

Artificial micromotors have gained an ever-growing interest, in recent years, as biomimetic devices and for their manifold microfluidic applications~\cite{Sanchez2015c}. 
Being able to convert ambient energy into autonomous motion, they fall in the realm of active matter~\cite{Ramaswamy2010a}.\\
A paradigmatic example of model micromotors is provided by self-phoretic particles (SPPs), which self-propel exploiting the phenomenon of colloidal phoresis~\cite{Anderson1989} in the inhomogeneous solute
distribution generated by a chemical reaction or a phase transition locally occurring at their surfaces.
SPPs include Janus metallic rods~\cite{Paxton2004}, 
catalytic~\cite{Howse2007,Theurkauff2012,Sanchez2015c} and light-activated Janus
colloids~\cite{Volpe2011,Palacci2013}, among others 
(see also~\cite{Paxton2006,Ebbens2010,Bechinger2016b} and references therein for reviews). Extensively
studied are suspensions of platinum coated, 
micron-sized polymeric spheres in an 
aqueous solution of hydrogen peroxide; the latter undergoes 
a decomposition reaction, catalyzed by the platinum, into
water and oxygen. For this reason, in what follows we will often refer to the solute as oxygen; however, as it will be made clear, 
the model simply requires the generation and 
diffusion of a scalar field. The approach is, therefore, more 
general and can simulate SPPs based on different mechanisms,
like the critical water-lutidine demixing~\cite{Buttinoni2013}, 
or even self-thermophoretic colloids~\cite{Golestanian2012,Ripoll2014}.\\
Despite the consistent body of theoretical, computational and experimental works witnessed, many questions remain still unanswered in the physics of SPPs, especially when it 
comes to more complex environments than the bulk of a fluid, as, e.g., in the presence of interfaces~\cite{Bechinger2016b}.
%For instance, although these active particles move inside a fluid, the media where its natural analogues move is complex, and can be full of interfaces\cite{Bechinger2016b}. 
As a matter of fact, how the motion of these active particles could be modified or steered by the presence of an interface is not yet totally understood. 
Recent studies have focused on solid-liquid interfaces\cite{Simmchen2016b, Katuri2018c,Uspal2015a}, liquid-liquid interfaces\cite{Dominguez2016a,Malgaretti2016,Wang2016,Peter2020a} or even on a combination of both\cite{Palacios2019b}, but a theoretical/computational 
approach taking into account hydrodynamics and a thermodynamically consistent modelling of the 
multiphase solute-solvent-particle system is so far missing.
Indeed, liquid-liquid interfaces add extra degrees of freedom to the system, owing to their deformability and to the solid phase wetting properties, thus significantly enriching the particle dynamics. 
Given also the intrinsic out-of-equilibrium physics of active particles, it appears clear how modelling may represent a challenging task.\\
%But due to the out-of-equilibrium physics that these particles %have, and since the rich dynamics of these interfaces, it is difficult to draw a simple model. 
%Most of these studies have focused on the static case of active particles at these interfaces, but little has %been said about letting free these particles to move while the interface is free to evolve.
In this paper, we propose a mesoscopic numerical model, based on a multiphase lattice Boltzmann method, featuring 
a free energy functional depending on two phase fields that describe the two immiscible component
mixture (e.g., oil and water) and the product (oxygen), respectively. The suspended solid particles are 
endowed with the capability of performing diffusiophoretic motion and of generating a solute field, which stems from the activity of the catalytic site. 
The method proves, then, able to handle at the same time solvent 
hydrodynamics,
particle-solute interactions, giving rise to the self-propulsion, wettability 
and preferential solubility of oxygen, that can, in general, 
accumulate more in one of the two liquid phases, thus 
allowing to simulate different combination of immiscible fluids.\\
The paper is organized as follows. In Sec.~\ref{sec:model} the 
thermohydrodynamic model is introduced and tested together with the description of the fluid-solid coupling and of the implementation 
of the self-diffusiophoresis. 
In Sec.~\ref{sec:results} we present the model validation against
controlled setups, starting with that of an isolated SPP
in a single phase fluid and then moving to the case of mixtures,
distinguishing between active and inactive 
particles, in order to disentangle the effects of 
capillary and phoretic forces. We report results showing 
that the relative
position and orientation of a self-phoretic Janus particle and an interface depend, in a non-trivial way, on 
wettability, phoretic mobilities and oxygen solubility.
Our findings suggest, then, that a proper tuning of such parameters may enable the guidance of active particles in
non-homogeneous fluid media. 
Conclusions and perspectives are drawn in the final 
Sec.~\ref{sec:conclusions}

\section{Computational model}\label{sec:model}
A suspension of active particles in presence of liquid-liquid interfaces consists of a fluid phase (solvents $+$ solute)
and a solid phase (the active particles).  
To model such a
multiphase (and multicomponent) system we resort to a mesoscopic approach, based on the lattice Boltzmann (LB) method
\cite{Benzi1992,Wolf-Gladrow2000,KruegerBook} in the 
phase field formulation \cite{Swift1996,KENDON2001a,Gonnella2019}.

\subsection{Phase field model: the free energy functional}
\begin{figure}[!hbpt]
%    \centering
    \includegraphics[scale=0.2]{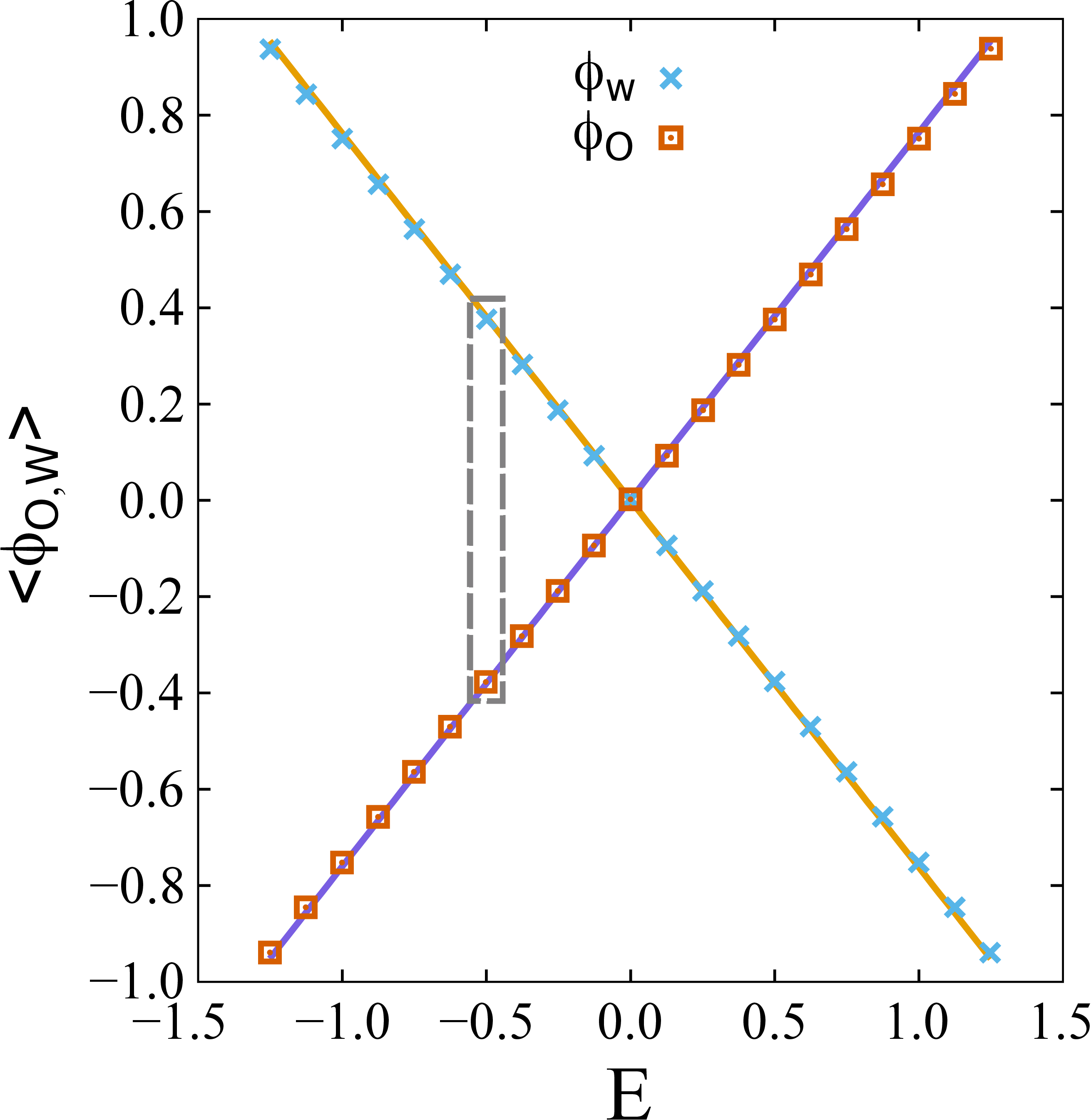}
    \caption{Equilibrium oxygen concentration, averaged respectively over the water (blue crosses) and oil (orange squares) phases, as a function of
    the $E$ parameter. The solid lines represent the expectations from the thermodynamic model, 
    $\langle \phi \rangle_{O,W} \approx \pm E \tanh(1)$, where the positive/negative sign corresponds to oil/water and is depicted in violet/yellow. The simulation were run with $\phi_{b0}=0$. }
    \label{Fig:Ludwig_order_param}
\end{figure}

The fluid phase is a ternary mixture made of two immiscible liquids (say, water and oil) and a solute, which is the product of 
the reaction occurring at the catalytic site on the particle surface (the oxygen). We associate to the water-oil system a scalar field $\psi(\mathbf{r},t)$ standing for the local 
composition, that is $\psi=\frac{\rho_W-\rho_O}{\rho_W + \rho_O}$, where $\rho_W$ and $\rho_O$ are the density fields of water and oil, respectively. As in the standard 
Cahn-Hilliard theory, the thermodynamics of the oil-water mixture is controlled by a quartic in $\psi$ double-well 
free energy density of Landau type, $f_{\text{\tiny{OW}}}[\psi] = \frac{A}{4}\psi^4 + \frac{B}{2}\psi^2$ (with $A>0$ and $B<0$). 
This free energy has to be extended to embrace the dynamics of the oxygen, that is, in principle, miscible with each of the two 
other components; therefore we need to add a term characterized by a single minimum that disregards the energetic cost associated to the concentration gradients \cite{Scagliarini2020a}, such that, in the case of a single component solvent ($f_{\text{\tiny{OW}}}=0$ identically),
a diffusive equation for the solute is recovered. A simple parabolic potential is appropriate to this aim (as we will show shortly), 
namely $f_{\text{\tiny{$O_2$}}}[\phi]=\frac{C}{2}\phi^2$ ($C>0$), 
having introduced the field $\phi(\mathbf{r},t)$. 
In actual systems, though, the oxygen may display, in general, a greater affinity for one of the two liquids (it can be more soluble in water than in oil, or vice versa). To account for this 
preferential concentration an "interaction" term, coupling $\phi$ and $\psi$, has to be included. We propose to do so by simply shifting 
the global minimum of 
$f^{(0)}_{\text{\tiny{$O_2$}}}$ in $\phi=0$ to a $\psi$-dependent minimum, i.e. 
$f^{(0)}_{\text{\tiny{$O_2$}}}[\phi]=\frac{C}{2}\phi^2 \rightarrow f_{\text{\tiny{$O_2$}}}[\phi,\psi] = \frac{C}{2}\left(\phi - \phi_0(\psi)\right)^2$.
For $\phi_0(\psi)$ we choose the form $\phi_0(\psi)=\phi_b + E \tanh(\psi)$, where the parameter $E$ tunes the oxygen solubility and $\phi_b$ sets the average 
oxygen concentration.
The full free energy functional then reads:
\begin{widetext}
\begin{equation}
  %  F[\phi, \psi] = \int \textbf{dr} \left[ \frac{A'}{2}(\phi-E \tanh(\psi)-{\phi_{b0}})^2 + \frac{A}{4}\psi^4 - \frac{A}{2}\psi^2 +\frac{\kappa}{2}(\partial_\alpha \psi)^2\right],
    F[\psi,\phi] = \int \textbf{dr} \left[\frac{A}{4}\psi^4 + \frac{B}{2}\psi^2 +\frac{\kappa}{2}|\nabla \psi|^2 + \frac{C}{2}(\phi-E \tanh(\psi)-{\phi_{b0}})^2 \right].
		\label{eq:free_energy}
\end{equation}
\end{widetext}
Hereafter we set $B=-A$, such that the $\psi$ minima, corresponding to the bulk water and oil phases, are located in $\psi = \pm 1$.  
%where $A', A, \kappa$ are constants that describe both components and $E$ is a constant that modulates the quantity of $\phi$ is in each phase of $\psi$. We also include $\phi_{b0}$ as a constant that %considers the initial global average $\phi$ in the simulated box. Thus, the term $\phi_0(\psi)= E\tanh(\psi)+\phi_{b0}$ accounts for a minimum of $\phi$ that is a function of the binary phase described by %$\psi$. Here the sign of these constants are positive for the case of $A', A, \kappa$, and positive or negative for $E$ and $\phi_{b0}$. We decided to model the minimum in terms of an hyperbolic tangent since %by the selection of our constants, $\psi \in [-1,1]$ and hence this function expresses a soft change with $\psi$. 
The minimization of the functional ($\ref{eq:free_energy}$) yields the chemical potentials $\mu_{\phi} = \frac{\delta F}{\delta \phi}$ and $\mu_{\psi} = \frac{\delta F}{\delta \psi}$:
\begin{widetext}
\begin{equation}
	\left.\begin{aligned}
   		\mu_\phi&=C(\phi-E \tanh(\psi)-\phi_{b0})\\
    	\mu_\psi&=A(\psi^3 - \psi) -\kappa\nabla^2 \psi - C \, E\frac{(\phi-E \tanh(\psi)-\phi_{b0})}{\cosh^2(\psi)} 
	\end{aligned}\right\}.
	\label{eq:chem_potentials}
\end{equation}
\end{widetext}
The dynamics of the ternary mixture system is, then, described by the following equations:
\begin{widetext}
\begin{equation}
	\left.\begin{aligned}
	   	\partial_t\phi + \nabla \cdot \left( \bm{u}\phi \right) &= D_\phi \nabla^2\left(\phi -\phi_0(\psi) \right) + \mathcal{D}_\phi-k_d(\phi-\phi_0(\psi))\\
    	\partial_t\psi + \nabla \cdot \left( \bm{u}\psi \right) &= D_\psi \nabla^2\left(\psi^3 - \psi -\frac{\kappa}{A}\nabla^2 \psi - E\frac{\phi-\phi_0(\psi)}{\cosh^2(\psi)} \right)
	\end{aligned}\right\},
	\label{eq:diffusion_reaction}
\end{equation}
\end{widetext}
where $D_\phi= C \, M_\phi$ and $D_\psi= A \, M_\psi$ are the diffusivities for the oxygen and the water-oil mixture, respectively, and $M_\phi$ and $M_\psi$ are the mobility constants 
for $\phi$ and $\psi$. The equation for $\phi$ has been equipped with a source term,  $\mathcal{D}_\phi$, that accounts for the generation of oxygen in a reaction catalyzed by the particles (see next section for further details). 
This production needs to be balanced by a sink term, $k_d(\phi-\phi_0(\psi))$, in order to allow the attainment of a steady state. 
Physically, the sink mimics the degradation of the production or its loss in the environment~\cite{Scagliarini2020a}.
%introduces a sink of the product to avoid a continuous increasing of $\phi$ in our system to avoid numerical instabilities as refer in Ref.~\cite{Scagliarini2020a}. 
%As seen in Fig.~\ref{Fig:Ludwig_order_param}, we can achieve a great control of the system when running a simulation without particles. 
Fig.~\ref{Fig:Ludwig_order_param} displays the effect of changing the solubility parameter from negative to positive values, by plotting the average oxygen concentration, 
$\langle \phi \rangle_{O,W}$ in oil or water as 
a function of $E$, at equilibrium and in the absence of particles (hence of oxygen production). 
Since, by virtue of Eq.~(\ref{eq:chem_potentials}), the equilibrium profile of $\phi$ is $\phi = E \tanh(\psi)$ 
(the background value having been set to zero here, $\phi_b=0$), the average, up to terms of infinitesimal order in $\xi/L$ ($\xi$ 
being the interface width and $L$ the system size), is $\langle \phi \rangle_{O,W} \approx \pm E \tanh(1)$, where the positive/negative 
corresponds to the average being taken over the oil or water phase, respectively; this prediction is reported in Fig.~\ref{Fig:Ludwig_order_param} with solid lines and agrees well with the numerical data.
%By selecting a proper value of the coupling term $E$, we can set a difference in $\phi$ in both phases of the binary mixture. Because the values of $\psi$ in both phases are set by the %thermodynamic value $A$, the equilibrium value for $\phi$ is linear with $E$ as already expressed if $\phi_{b0}$ is set to zero. Thus, we can see how this linearity also appears in %Fig.~\ref{Fig:Ludwig_order_param}. 

\subsection{Particles}

\begin{figure*}[!htpb]
    %\centering
    \includegraphics[width=\textwidth]{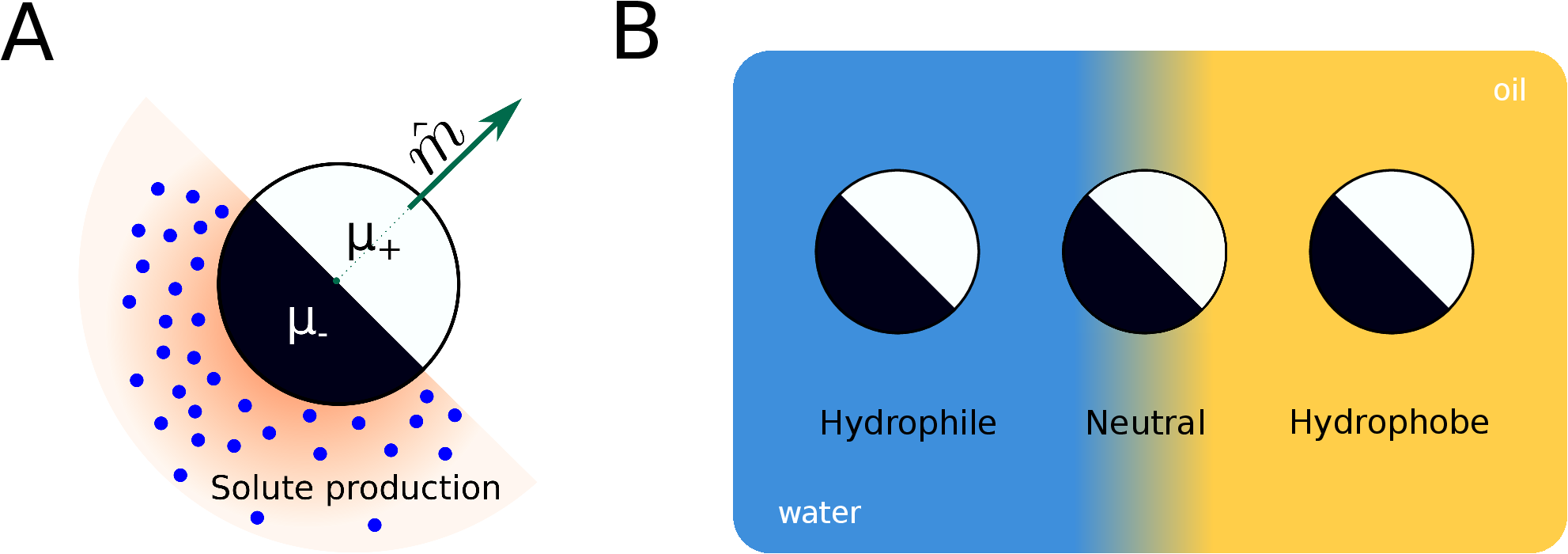}
    %\caption{{\bf A. Janus particle scheme.} We compute Janus particles as spherical particles that behave differently in both of their half sides. One side is active, meaning that produces a component, described by our order parameter $\phi$, that able them to self-propel by a self-diffusiophoretic motion. This surface side interacts with the product via a diffusiophoretic mobility $\mu_A$, which controls the slip speed over this side. On the other side, the inactive side does not produce this component, and react with the existing with the same or different value for the diffusiophoretic mobility, which in this case we named as $\mu_I$. In any case, the product created by this particle is removed at each step with a characteristic length L, which depends on a degradation ratio $\kappa$ and its diffusion constant $D$. {\bf B.} Our Janus particles will be in presence of a liquid-liquid interface. We include into them the effects of wetting, letting particles to naturally be in one phase of the binary mixture, in the other phase or in the interface. As a way to represent this binary mixture we can think in water and oil, and hence wetting gives us hydrophilic, hydrophobic or neutral behaviour.}
    \caption{{\bf A)} Sketch of a spherical self-diffusiophoretic particle. The particle is characterized by a Janus profile of both
    the activity (it produces solute only over one hemisphere) and the phoretic mobility, whose value is $\mu_-$ over the active cap and 
    $\mu_+$ otherwise. {\bf B)} Active particle in the presence of an interface: depending on the value of the contact angle, $\theta$, it tends to stay preferentially in the bulk of the water phase ($\theta>90^{\circ}$), of the oil phase ($\theta<90^{\circ}$) or absorbed at the interface ($\theta=90^{\circ}$).}
    \label{Fig:Ludwig_particles}
\end{figure*}

Particles are modelled as solid spheres defined by a set of boundary "links" between inner and outer nodes. 
The fluid-solid coupling is realized by means of the so called "bounce-back-on-links" algorithm that guarantees 
the proper momentum-torque exchange between particles and solvent \cite{Ladd1,Ladd2,Nguyen,Aidun}. The colloidal 
phoresis is introduced by imposing at the particle surface an effective slip velocity profile which depends on the 
local solute concentration~\cite{Anderson1989} as:
\begin{equation}
    \bm{v}_s=\mu(\bm{r}_s)(\bm{1}-\bm{\hat{n}}\times\bm{\hat{n}})\cdot\nabla \phi,
\end{equation}
where $\bm{r}_s$ is a point on the surface of the particle, $\bm{\hat{n}}(\bm{r}_s)$ is the normal to the surface in $\bm{r}_s$ and $\mu(\bm{r}_s)$ is the phoretic mobility at $\bm{r}_s$, which carries the molecular details of the solute-colloid interaction\cite{Anderson1989}.
As a consequence, in the presence of concentration gradients, particles gain a net propulsion velocity 
$\bm{V}_p \sim -\mu\nabla \phi$ (for uniform phoretic mobility $\mu(\bm{r}_s) \equiv \mu$), hence if $\mu<0$ they
are attracted by the solute, else if $\mu>0$ they are repelled.
To achieve self-propulsion, particles are, then, endowed with the property of generating solute~\cite{Golestanian2007c};
this is done by simply adding a production term that injects $\phi$ with a given rate at nodes neighbouring the particles
surfaces, thus modelling the catalytic activity of $Pt$-coated colloids. 
In particular, a Janus activity profile is chosen:
\begin{equation}\label{eq:activity}
    \Pi(\bm{r}_s)=
      \begin{cases}
        \alpha & \text{if $\bm{\hat{m}}\cdot\bm{\hat{n}}(\bm{r}_s)\leq0$ } \\
        0 & \text{if $\bm{\hat{m}}\cdot\bm{\hat{n}}(\bm{r}_s)>0$}
      \end{cases},    
\end{equation}
where $\alpha$ is the constant production rate and $\bm{\hat{m}}$ is the particle characteristic unit vector (see the sketch in Fig. \ref{Fig:Ludwig_particles}A)). Notice that the superposition of such activities associated to the various particles is precisely 
what gives rise to the production term $\mathcal{D}_{\phi}$ appearing in Eq.~(\ref{eq:diffusion_reaction}). 
Analogously, for the phoretic mobility $\mu(\bm{r}_s)$ we set:
\begin{equation}\label{eq:MuJ}
    \mu(\bm{r}_s)=
      \begin{cases}
        \mu_- & \text{if $\bm{\hat{m}}\cdot\bm{\hat{n}}(\bm{r}_s)\leq0$ } \\
        \mu_+ & \text{if $\bm{\hat{m}}\cdot\bm{\hat{n}}(\bm{r}_s)>0$}
      \end{cases}.    
\end{equation}
For an isolated Janus particle with the above activity and mobility profiles we expect a motion with constant velocity
of magnitude~\cite{Golestanian2005c,Golestanian2007c,Popescu2010c}
\begin{equation}\label{eq:Vp}
v_p=\frac{|(\mu_+ + \mu_-)|\alpha}{8D}.
\end{equation}
In the presence of interfaces a specific treatment of the interaction of the two liquids with solid boundaries, which determines 
the particles wetting properties, needs to be included. To this aim, an extra boundary term is added to the free energy functional, 
such that:
\begin{equation}\label{eq:fe_wetting}
    F_{\text{\tiny{tot}}}[\phi, \psi]=F[\phi, \psi] + \int_S H \psi(\bm{r}_s) d\bm{r}_s
\end{equation}
where the integral is over the solid surface. The parameter $H$ controls the wetting through the following boundary condition, that 
can be derived by minimization of the surface term in (\ref{eq:fe_wetting}), 
\begin{equation}
    H=\kappa|\nabla\psi\cdot\bm{\hat{n}}|
\end{equation}
and, therefore, it sets the particle contact angle $\theta_c$, to which it is related by~\cite{Desplat2001b}
\begin{equation}
    \cos(\theta_c) = \frac{1}{2}\left[-(1-h)^{3/2}+(1+h)^{3/2}\right]
\end{equation}
\noindent where $h=H\sqrt{1/(\kappa A)}$. 
%As a consequence, each of the active particles can feel a different wetting property, which will modify the capillarity forces that %the colloid has with the interface as seen in Fig.~\ref{Fig:Ludwig_particles}B. This allows  to study the competition between activity %and capillarity close to these interfaces for active particles. \\
%Finally, Janus particles  include a soft-potential for close-to-contact particles encounters, as well as  the possibility to add %gravity following Ref.~\cite{Scagliarini2020a}.

\subsection{Numerical details}

We simulate numerically the model just introduced on three-dimensional periodic 
lattices of sizes ranging between $32^3$ to $64 \times 96 \times 96$ (with unit 
spacing, $\Delta x = 1$).
The two liquids have the same kinematic viscosity, equal to $\nu=1/6$, 
and same density $\rho =1$, in lattice Boltzmann units ($\text{lbu}$).
The free energy parameters are set to $A=C=0.0625$, $\kappa=0.04$, 
such that the surface tension is $\sigma=(2/3)\sqrt{2\kappa A}\approx 0.047$,
and the mobilities are $M_{\psi}=0.4$ and $M_{\phi}=0.8$, giving the diffusion 
coefficients $D_{\psi}=0.025$ and $D_{\phi}=0.05$.
The particle radius is fixed to $R=4.5$. The activity is varied in the range
$\alpha \in [0, 10^{-2}]$ and the phoretic mobilities in 
$\mu_{\pm} \in [0, 0.6]$ (we consider only oxyrepulsive particles).
Correspondingly, the largest Reynolds, Mach and P\'eclet numbers are of the order
$Re \approx 0.1$, $Ma \approx 0.03$ and $Pe \approx 3$ (although in most of the simulations we have $Pe \approx 0.1$), 
thus we are legitimately in an incompressible creeping flow regime.\\
Unless otherwise specified, the system is initialized with two slabs of oil and 
water separated by a flat interface (actually two, due to the periodic boundary
conditions), corresponding to the equilibrium hyperbolic tangent profile
$\psi(\mathbf{r},0) = \tanh(x/\xi)$.
The oxygen field is initially set to $\phi(\mathbf{r},0)=0$, everywhere, and 
then let equilibrate. 

\section{Results and discussion}\label{sec:results}

\subsection{Motion of a Janus particle in a single phase fluid}
\begin{figure*}[!htpb]
    %\centering
    \includegraphics[width=\textwidth]{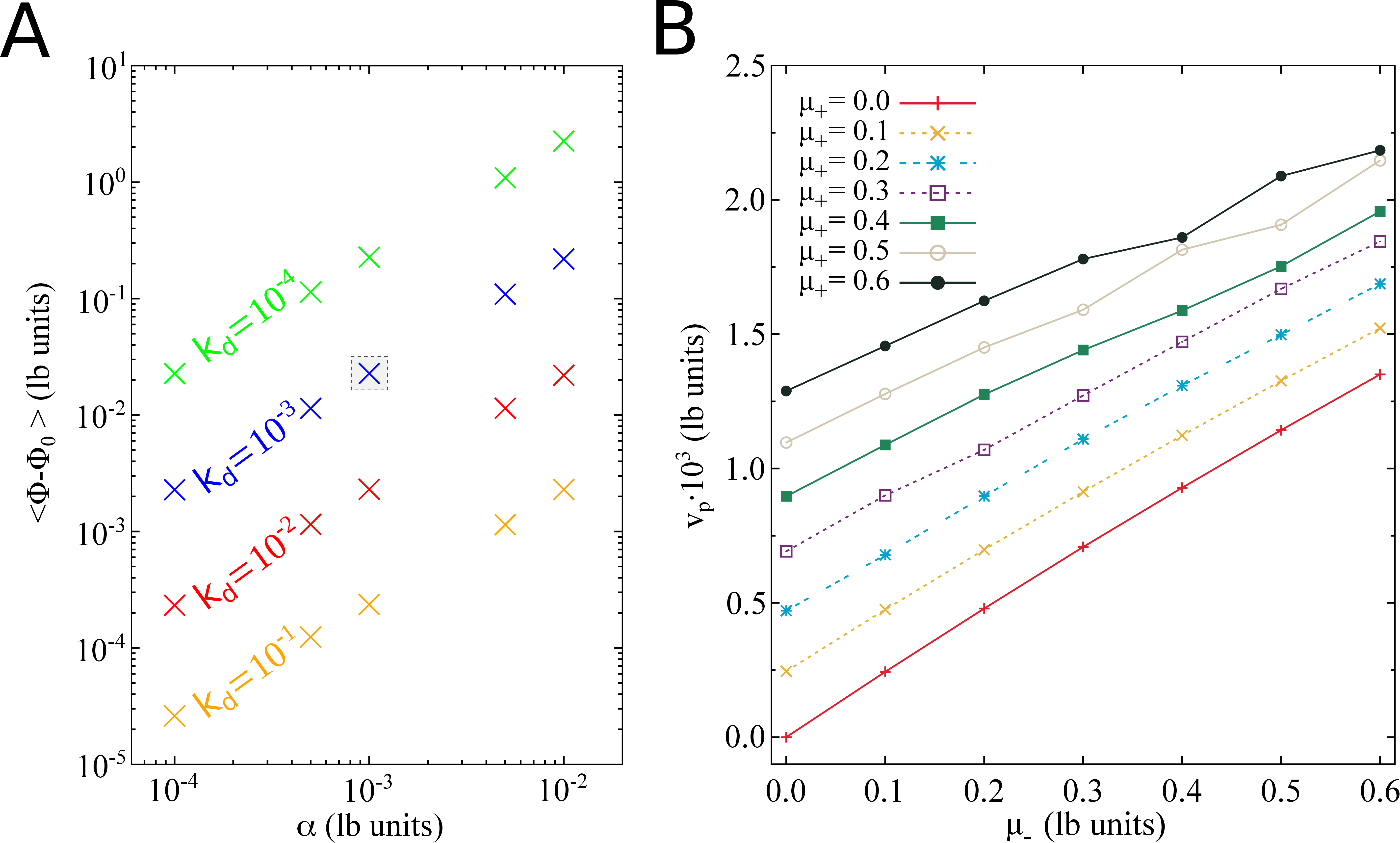}
    \caption{Motion of an active Janus particle in a single phase fluid. \textbf{A)} Steady state value of the space-averaged increase (with 
    respect to the initial time) of oxygen concentration,
    $\langle (\phi - \phi_0) \rangle$, as a function of the production rate 
    $\alpha$ (in $\text{lbu}$), for various degradation rates, $k_d$. The rest 
    of the data shown in the paper were obtained from simulations with $\alpha=10^{-3}$ and $k_d=10^{-3}$.
    \textbf{B)} Particle speed as a function of the phoretic mobility $\mu_-$, 
    for different values of $\mu_+$ (see Eq.~(\ref{eq:MuJ})), showing a linear 
    dependence as predicted by Eq.~(\ref{eq:Vp}).}
    \label{Fig:Ludwig_degradation_speed}
\end{figure*}

\noindent As a validation of the model, we first consider the motion of an isolated active particle in the bulk of a single phase fluid 
($A=E=\kappa=0$ and $\psi=0$, identically). 
%The motion of Janus particles will depend on the media properties, characterized %by the thermodynamic constants for the medium described in table %\ref{Table:Ludwig_parameters} but also, on several colloidal properties. These %properties are the ratio of creation of the product per node in one face of the %particle, $\alpha$, and the surface mobilities $\mu_A$ and $\mu_I$. The more %product they create by time step, the fastest the product will accumulate in the %media. Hence, a degradation of $\phi$ is also introduced, where for each time step %and node  its $\phi$ value is checked and  a quantity proportional to its distance %from equilibrium ($\phi-\phi_0$) is removed. This degradation is modulated by a %proportional constant $k_d$. Fig.~\ref{Fig:Ludwig_degradation_speed}A) displays %the change of the global average of  $\phi$  when considering   a Janus particle %in a single fluid phase, as a function of the activity and degradation rate. The %steady state values are used to  identify reference parameters for the average %$\phi$ and $k_d$. 
%We take this data after a short period of time where particles reach a stable %speed and  $\phi$ reaches a global constant average. From this map of quantities, %we decided to fix all our next simulations on a couple of $\phi$ and $k_d$. 
Fig.~\ref{Fig:Ludwig_degradation_speed}A reports the results of a set
of simulations 
aimed at tuning the degradation rate $k_d$. We notice, first of all,
from the plot of the space-averaged oxygen concentration, 
$\langle \phi \rangle (t)$, vs time (in the inset),
that the introduction of the sink term works as expected and a steady state 
is reached. The stationary value, $\langle \phi \rangle_{\infty}$, will depend,
of course, on both the particle activity and the degradation rate, as 
shown in the main panel of the same figure. In particular, it grows with $\alpha$
and decreases with $k_d$. The presence of a linear degradation term implies 
that the concentration field does not decay purely algebraically with the distance
from the source (the particle surface) but it is modulated by an exponential
factor, $\phi(r) \sim e^{-r/\ell}/r$, with screening length 
$\ell = \sqrt{D/k_d}$. 
In the remainder of the paper the value of the degradation rate is kept 
fixed to $k_d=10^{-3}$, which gives a screening length of approximately 
one particle diameter, $\ell \approx 2R$.\\
In Fig.~\ref{Fig:Ludwig_degradation_speed}B we check the dependence of the particle speed on the phoretic mobilities, at fixed activity $\alpha=10^{-3}$,  
plotting $v_p$ vs $\mu_-$ (the mobility value on the active side) for various values of the mobility on the opposite cap, $\mu_+$. 
As expected from the theoretical prediction, Eq.~(\ref{eq:Vp}), the speed 
grows linearly with $\mu_-$ (and $\mu_+$). The linearity deteriorates a bit 
as $\mu_{\pm}$ increase, probably due to the fact that the P\'eclet number 
is also increasing and tends to approach unity (we recall that the result 
(\ref{eq:Vp}) is derived under the assumption of vanishing $
Pe$~\cite{Golestanian2005c,Golestanian2007c}).
%If both mobilities are zero, the particle cannot interact with the product and %hence their speed vanish. However, once the surface mobility is not zero, even if %the mobility for the active side is zero, the particle will interact with the %product and will start to move. Because we focus in the low P\'eclet number  %regime, the increase of these mobilities causes a linear increase with the %particle speed~\cite{Golestanian2005c,Golestanian2007c,Scagliarini2020a}.

\subsection{Inactive Janus particles at liquid-liquid interfaces}
\begin{figure*}[!t]
	%\centering
	\includegraphics[width=\textwidth]{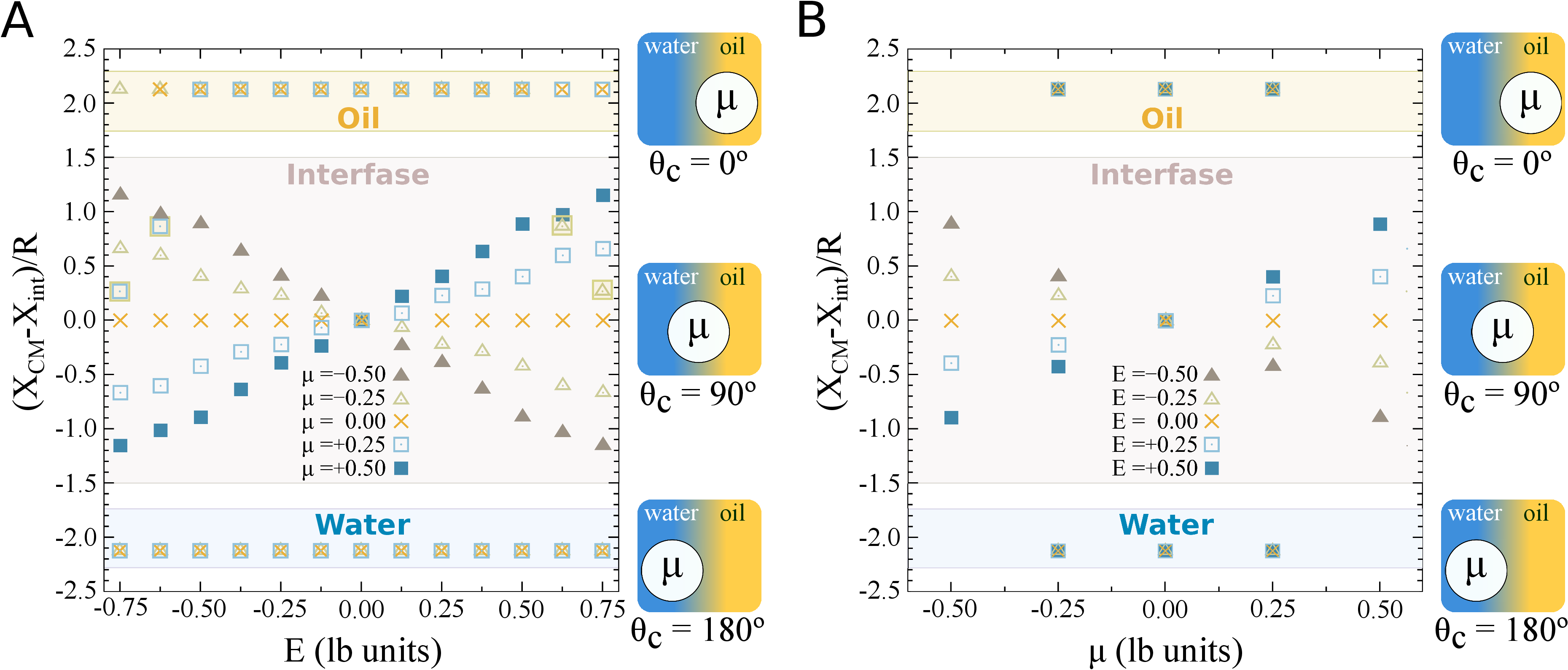}
	\caption{Equilibrium position, relative to the interface, of an inactive ($\alpha=0$) particle with uniform phoretic mobility, as a function
	of $E$ for various $\mu$'s (panel A) and as a function of $\mu$ for various $E$'s (panel B). 
	Data for three different contact angles are shown; particles are initially 
	placed in oil (hydrophobic, $\theta_c=0^{\circ}$), water (hydrophilic, $\theta_c=180^{\circ}$ or at the interface 
	(neutral, $\theta_c=90^{\circ}$).}
	\label{Fig:Ludwig_inactive_equal}
\end{figure*}

\noindent Before facing the problem of active particle motion, a needed 
preliminary step is to investigate the interaction of an inactive 
diffusiophoretic particle with the interface, in order to analyze the competing effect of capillary and phoretic forces.
To this aim, first we focus on a particle with $\alpha=0$ and uniform phoretic mobility, $\mu_+=\mu_-=\mu$, initially placed either in oil, water or at
the interface, depending on whether it is hydrophobic ($\theta_c=0^{\circ}$), neutral ($\theta_c=90^{\circ}$) and hydrophilic ($\theta_c=180^{\circ}$).
Because of the imbalance of capillary and phoretic forces, the particle will relax from its initial position towards or away from the interface.
We then monitor its equilibrium position relative to the interface,  $(X_{\text{CM}}-X_{\text{int}})/R$,
as a function of $\mu$ and the oxygen solubility parameter $E$.
%To modulate the diffusiophoretic forces  simulations are run as a function of  %$\mu$ and $E$. Capillarity forces are modulated by changing the wetting of the %particle. We only study those wetting values that are extreme (totally %hydrophobic, totally hydrophilic, and neutral), and  place the particle in the %oil, water, or interface region respectively. Because of the imbalance of %capillary forces, particles will  relax from their initial position toward or away %from the interface. Therefore, we observe at which position particles move because %of the previous forces. Figs.~\ref{Fig:Ludwig_inactive_equal}A and %\ref{Fig:Ludwig_inactive_equal}B display the change in equilibrium position of the %Janus particle as a function of $E$ and $\mu$, respectively. \\
The results are shown in Fig.~\ref{Fig:Ludwig_inactive_equal}.
When particles are placed in the bulk of the oil or water phases, being the surrounding solute homogeneous, diffusiophoretic forces vanish and $\mu$ and $E$ do not affect the particle motion.\\
Conversely, particles initially trapped at the interface are surrounded by an inhomogeneous solute field, and diffusiophoretic forces become relevant. 
%To create the inhomogeneous field, $E$ is modulated. For $E=0$, $\psi$ is %homogeneously distributed across the fluid phases homogenous solute field, %implying that wetting determines the motion and equilibrium location of the Janus %particle at the liquid interface. A non-vanishing $E$ induces an asymmetric solute %field, which saturates to different values in the two coexisting liquid phases.
In particular, the larger is the difference of solute concentration in the two phases (i.e., for growing $|E|$), the stronger are these forces and the further they push the particle away from the interface. At the same time, phoretic forces depend on the strength of the particle-solute interaction, therefore increasing $|\mu|$ has the same effect as increasing $|E|$. 
More formally, at mechanical equilibrium phoretic and capillary forces balance each other along the normal to the interface, that is $F_{\text{cap}}=F_{\text{ph}}$.
The phoretic force is proportional to the concentration gradient, 
$F_{\text{ph}} \sim -\mu \nabla \phi$; next to the interface, we can 
approximate $\phi$ with a linear profile, by vertue of 
$\phi = E\tanh(\psi) = E\tanh(-\tanh(x/\xi)) \approx -2Ex/\xi$, such that 
the force reads, $F_{\text{ph}} \approx 2\mu E/\xi \propto \mu E$. 
For small interface deformations, capillarity acts as a Hookean restoring 
force, with an effective elastic constant proportional to the surface 
tension~\cite{JoannyDeGennes}, 
i.e. $F_{\text{cap}} \propto \sigma \Delta X$, whence
\begin{equation}
    \Delta X \equiv X_{\text{CM}} - X_{\text{int}} \propto \mu E,
\end{equation}
which explains the behaviour emerging from Fig.~\ref{Fig:Ludwig_inactive_equal}.\\
\begin{figure*}[!t]
	%\centering
	\includegraphics[width=\textwidth]{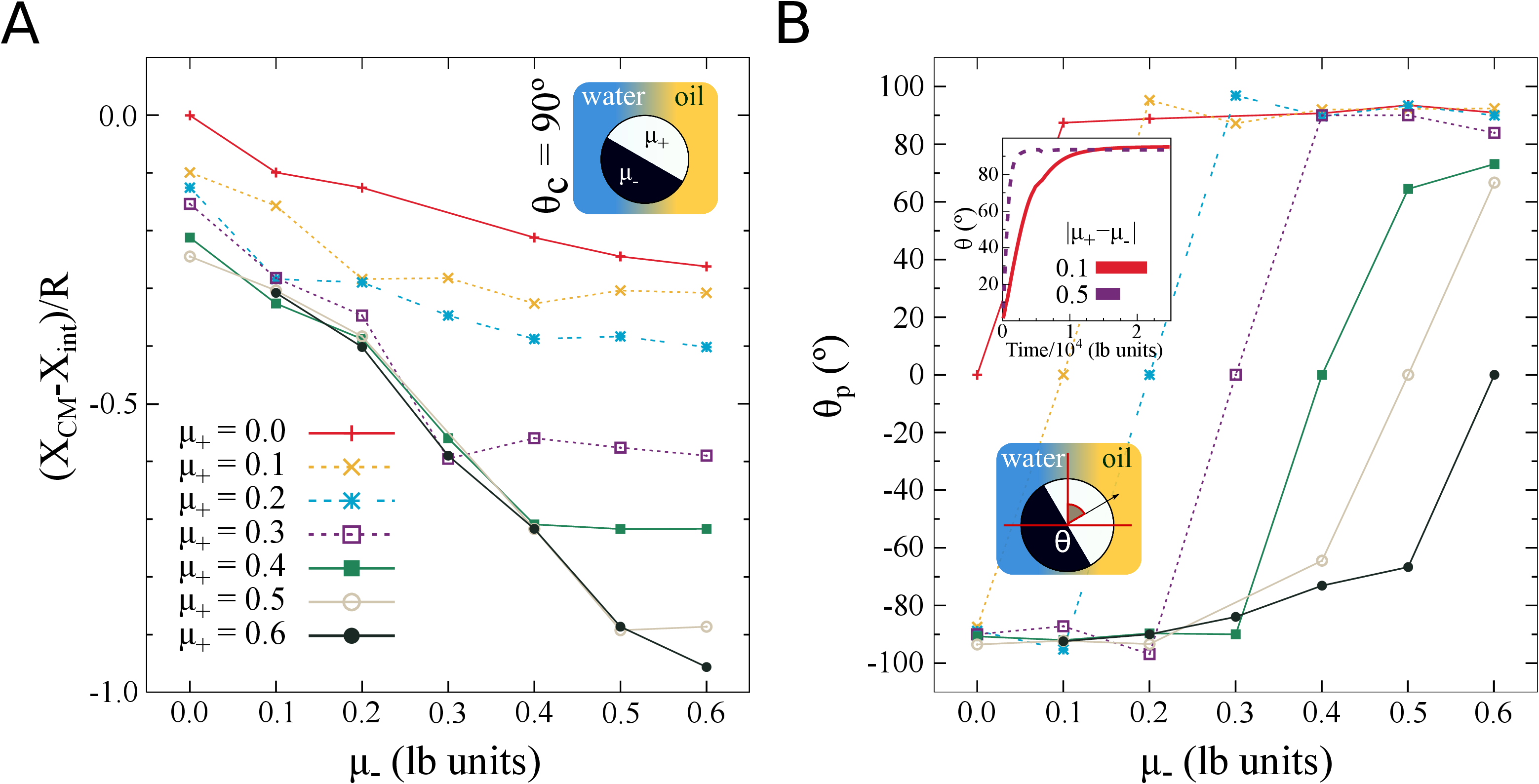}
	\caption{Equilibrium position (panel A) and orientation angle (panel B) relative to the interface of an inactive particle ($\alpha=0$) with 
	a Janus phoretic mobility profile ($\mu_+ \neq \mu_-$) and neutral wetting ($\theta_c = 90^{\circ}$) as a function of the rear mobility $\mu_-$, for various values of the front mobility $\mu_+$. The particle, initially placed at the interface with the characteristic vector $\hat{m}$
	in the interface plane, tends to escape from the plane by effect of the phoretic repulsion and to rotate due to the mismatching mobilities 
	inducing a torque (in the inset of panel B a typical time evolution of the orientation angle).}
	\label{Fig:Ludwig_inactive_inequal}
\end{figure*}
We next consider the case of inhomogeneous phoretic mobilities, $\mu_-\neq\mu_+$, when an oxygen concentration gradient is present at the interface, $E \neq 0$. 
We set $E=-0.5$, that leads to a larger oxygen concentration in the oil phase,
and impose neutral wetting ($\theta_c=90^{\circ}$). 
The particle is initially placed at the interface and aligned with it, i.e. 
its characteristic vector $\hat{m}$ lies in the interface plane and it is, then,
orthogonal to the concentration gradient, $\hat{m} \perp \nabla \phi$.
Consequently, due to the phoretic mobility mismatch, the particle is subject 
to a torque. We will consider here, therefore, both the 
equilibrium displacement and the equilibrium orientation angle, $\theta$,
with respect to the interface, as functions of $\mu_-$ for different
$\mu_+$ values.\\
%Although diffusiophoretic forces also depend on the magnitude of $E$, because the %larger the asymmetry in $\psi$ the stronger the effect, we focus on the impact %that $\mu$ has on particle motion, and fix $E=-1/2$ that leads to a larger %concentration of the solute in the oil phase (see %Fig.~\ref{Fig:Ludwig_inactive_equal}).
%Fig.~\ref{Fig:Ludwig_inactive_inequal}A shows that the  Janus particle,initially %aligned with its director vector $\hat{m}$ parallel to the interface, moves %further away from the interface as the diffusiophoretic mobilities increase their %magnitude.
As expected, Fig.~\ref{Fig:Ludwig_inactive_inequal}A shows that the particle relaxes to a position progressively further from the interface as the phoretic mobilities are increased. Interestingly, though, the equilibrium position
saturates at a finite distance from the interface when 
$\mu_- > \mu_+>0$.
%the value of the mobility in the leading side of the particle is larger than the %one in the rear part, except for $\mu_{down}=0$.
These observations are better understood looking at Fig.~\ref{Fig:Ludwig_inactive_inequal}B, where the equilibrium orientation angle
$\theta$ is plotted. 
%Initially the Janus particles are always facing the interface with larger mobility %(more repulsive) toward the water region (region with less solute). 
The phoretic torque induced rotation undergone by the particle is faster if the difference between both mobilities is larger, as expected (see inset of Fig.~\ref{Fig:Ludwig_inactive_inequal}B). Because particles reorient fast with the stronger phoretic mobility facing the water region, the side facing the oil is the more important input to displace the particle from the interface. Hence, this explains why in Fig.~\ref{Fig:Ludwig_inactive_inequal}A we reach a saturation when $\mu_- > \mu_+$.

\subsection{Active Janus particles and liquid-liquid interfaces}
\begin{figure*}[!htpb]
    %\centering
    \includegraphics[width=\textwidth]{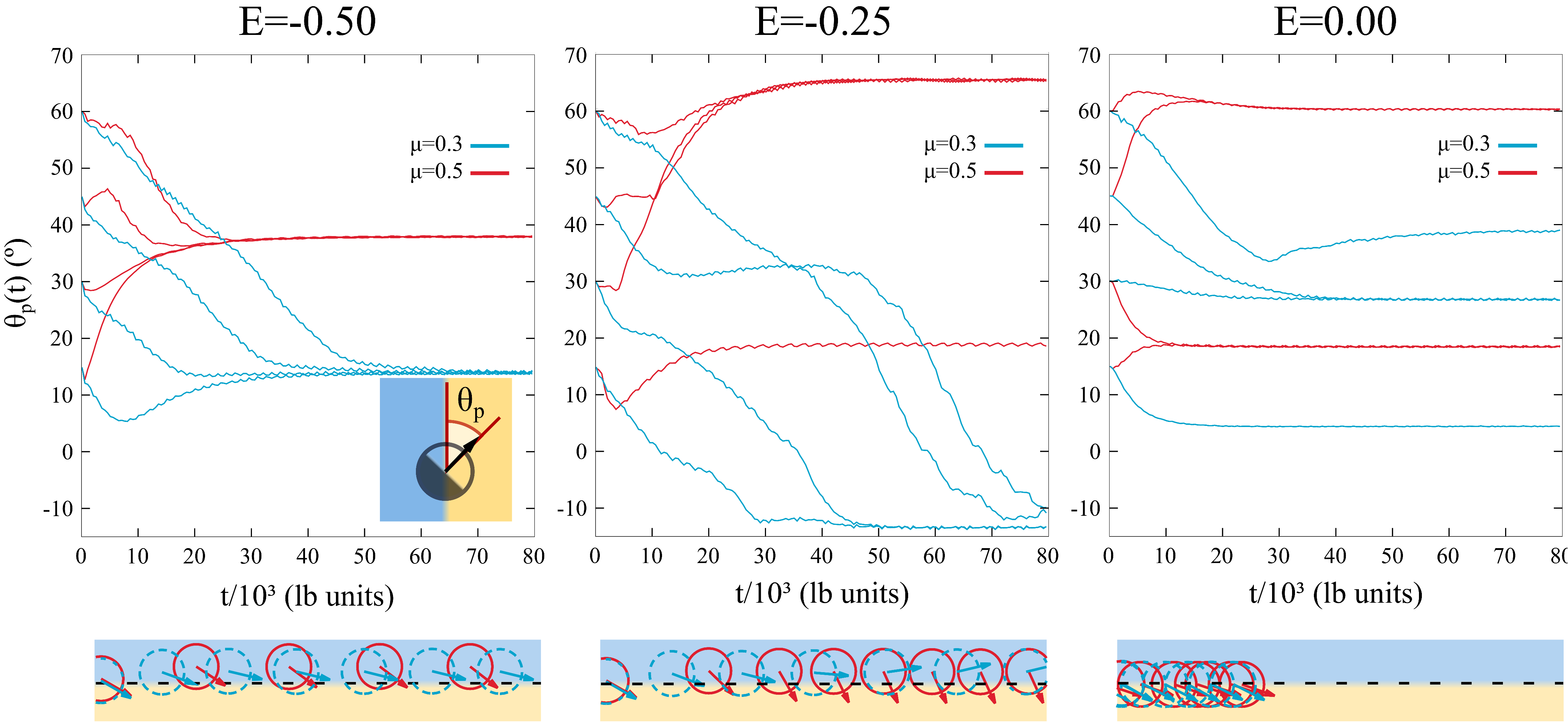}
    \caption{Angle between the particle characteristic vector and the interface, $\theta_p$, as a function of time for an active colloid
    with $\theta_c = 90^{\circ}$ (neutral wetting), trapped at the interface. The data sets correspond to various initial 
    orientation angles, two different surface mobilities ($\mu=0.3$ and $\mu=0.5$) and to three different oxygen-oil/water affinities: $E=-0.5$ (left panel), $E=-0.25$ (middle panel) and $E=0$ (right panel). The insets below the curves indicate the progression of the particles along the interface for the case with initial $\theta_p=60^{\circ}$ for the $8 \times 10^4$ time steps simulated.}
    \label{Fig:Ludwig_at-interface}
\end{figure*}

\begin{figure*}[!htpb]
    %\centering
    \includegraphics[width=\textwidth]{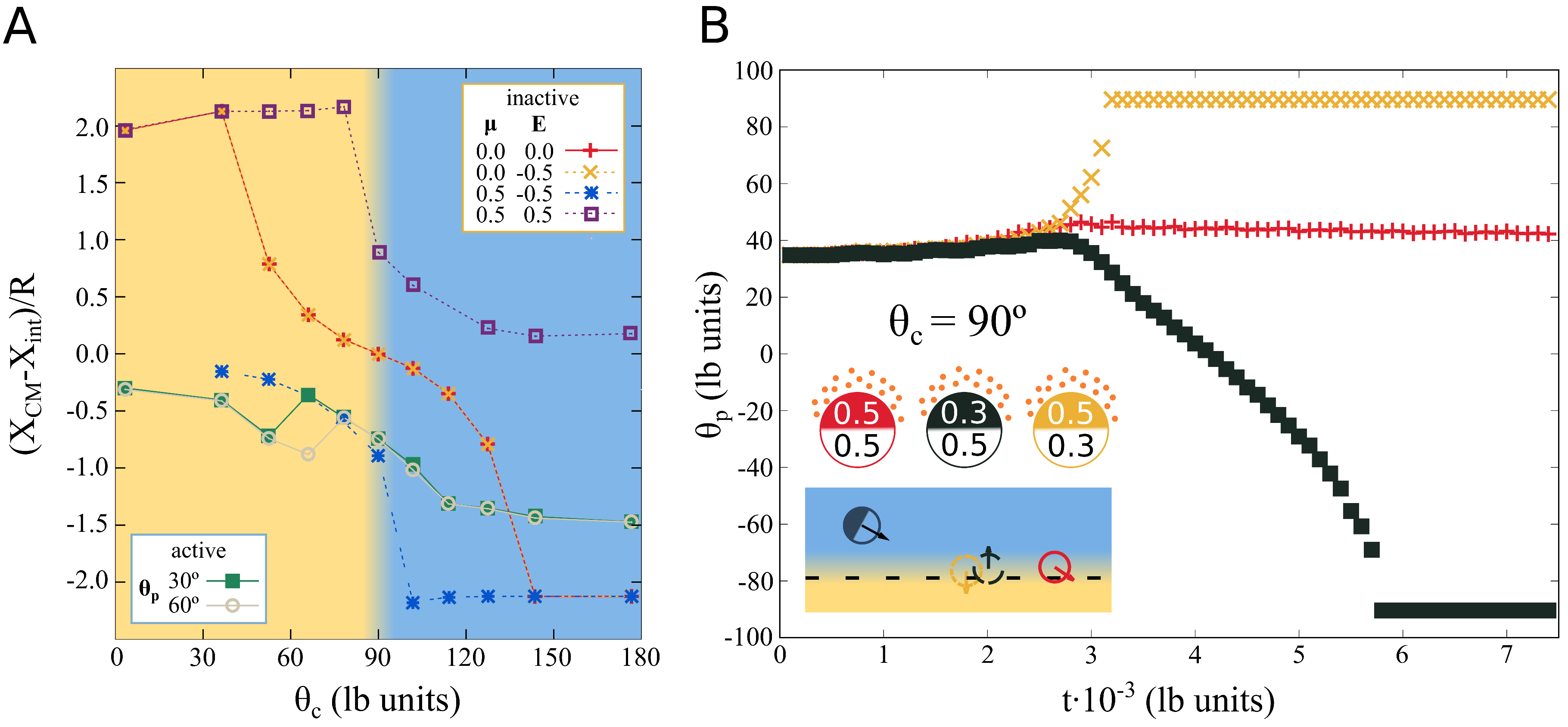}
    \caption{A) Steady particle displacement with respect to the interface as a function of the contact angle, for various combinations of $\mu$, $E$ and $\alpha=0$ (inactive) or $\alpha=10^{-3}$ (active). In the active case, data for two different values of the initial orientation angle are shown: $\theta_p^{(0)}=30^{\circ}$ (\color{green}{$\blacksquare$}\color{black}) and $\theta_p^{(0)}=60^{\circ}$ (\color{gray}{$\circ$}\color{black}).
    %Red and orange lines are for inactive particles without phoretic forces, while blue and violet have phoretic forces. Green and gray lines are %for active particles, with the difference of the initial angle condition ($\theta_p=60$º for gray and $\theta_p=30$º for green). 
    B) Trajectory (in a plane orthogonal to the interface) of an active particle initially placed in water and oriented towards the interface (indicated with the dashed line); the initial orientation angle is $\theta=30^{\circ}$ and the colloid is neutral 
    ($\theta_c =90^{\circ}$). Three combinations of phoretic mobilities, $(\mu_+,\mu_-)$, are considered: $(0.5,0.5)$ (red), $(0.5,0.3)$ (blue) and $(0.3,0.5)$ (green).}
    \label{Fig:Ludwig_interface-bounce}
\end{figure*}

Once the inherent behaviour of capillarity and wetting properties on passive Janus colloids has been established, we focus on the behavior of active Janus colloids, $\alpha \neq 0$, with a uniform phoretic mobility profile 
($\mu_- = \mu_+ \equiv \mu$). Initially, we consider neutrally-wetting particles trapped at the interface, and analyze their motion at varying $\mu$ 
and the oxygen solubility parameter $E$, quantifying the product concentration ratio at the two sides of the interface. The values $\mu=0.3$ and $\mu=0.5$, and  
$E=-0.5$, $E=-0.25$, $E=0$ (corresponding to no concentration mismatch) 
are used. We run simulations starting with different particle orientations defined by $\theta_p$, which is the angle between the particle characteristic vector and the interface (See Fig.~\ref{Fig:Ludwig_at-interface}). \\
When the difference of product in the two phases is high ($E=-0.5$) Janus particles move along the interface. Interestingly, if the simulations are initialized with different  $\theta_p$, particles stabilize at a unique angle, which depends on the particle surface mobility. When the ratio of products in both phases is closer to 1 (e.g. $E=-1/4$), particles continue their motion at interfaces, although they are slower, and a unique angle is no longer observed. For some mobilities, as in $\mu=0.3$, a single  angle is observed, but for others, as for $\mu=0.5$ we observe the appearance of competing attractors. 
%Moreover,  the asymptotic  $\theta_p$  changed from the previous case. 
Thus, both the asymmetric accumulation of the product in both sides and the surface mobility of the particle change the torque the particle feels at the interface and that stabilizes at a certain $\theta_p$. Finally, in the last scenario where both phases are symmetric with respect to the solute solubility, $E=0$,  particles move very slowly along the interface, and additional attractors  for $\theta_p$ appear. Consequently, the asymmetric accumulation of product is also responsible for the particles speed  at the interface.\\
We next analyze the impact of wetting, when a particle is initially placed at the interface and moves along it.
We consider both active and inactive particles and monitor the steady displacement of the particle from the interface plane, at changing the contact angle, the phoretic mobilities and the oxygen solubility parameter. The results are reported in Fig.\ref{Fig:Ludwig_interface-bounce}A.\\ 
%in trapped particles when phoretic forces compete with capillarity forces. To quantify such a competition,  particles are placed at a prescribed %position and, as shown in  Fig.\ref{Fig:Ludwig_interface-bounce}A,  the position at which they  stabilize is displayed.\\
Inactive, non-phoretic ($\mu=0$) particles are, of course, insensitive to variations of the oxygen field configuration and, hence, to $E$; 
this is reflected in the full overlap of the data for $E=0$ (\textcolor{red}{$+$}) and $E=-1/2$ (\textcolor{orange}{$\times$}).
In both cases, though, as expected for passive colloids, the more the contact angle departs from $90^{\circ}$, the further the particle settles away from the interface.\\
Remarkably, instead, for finite, large enough, phoretic mobility ($\mu=0.5$), phoretic repulsion is capable to overcome interfacial forces
and the particle tends to stay away from the phase richer in oxygen, be it water, $E>0$ (\textcolor{magenta}{$\square$}), 
or oil, $E<0$ (\textcolor{blue}{$\ast$}), irrespective of its wettability.
More precisely, taking, for instance, the case $E<0$, if phoretic and capillary forces have opposite directions ($\theta_c<90^{\circ}$), 
the particle can be stabilized, roughly, at the interface, but if they have the point towards the same side, the particle leaves 
the interface plane and there is, basically, no dependence on the contact angle.\\
%We, then, activate diffusiophoretic forces ($\mu=1/2$) and test at different $E$ fields ($E=-1/2$, blue line and $E=1/2$, violet lines). Particles %with repulsive (positive) mobilities (blue line, $E<0$) move towards low product concentration regions even if their wetting prefers regions of high %product concentration ($\theta_c<90$º) because diffusiophoretic forces overcome wetting forces. When both forces have opposite directions %($\theta_c<90$º), particles can stabilize at the interface instead of being in one side of the interface. But if forces have the same direction %($\theta_c>90$º), particles leave the interface, and there is no dependence on the wetting force. This effect also occurs in the opposite direction %if we make attractive (negative) the mobility or if we maintain the repulsion of the particle, but we change which phase concentrate more solute %(violet line). In any case, the addition of diffusiophoretic forces make the particles  move away from the phase with higher solute concentration %(oil phase, yellow area).\\
Active particles ($\alpha \neq 0$), for which only the case $E<0$ is shown, manifest a similar behaviour. However, the activity introduces 
an extra force which has a component normal to the interface (for the steady orientation angle differs, in general, from zero) and pushes the particle closer to the oxygen-rich region.
%Finally, we make particles active (green and grey lines). Active particles cannot enter into areas of high product concentration neither, even  if %particles move perpendicular to the interface. However, for hydrophilic particles ($\theta_c>90$º)  activity approaches them to the interface. 
To check if the initial orientation $\theta_p^{(0)}$ plays a role, simulations are run with $\theta_p^{(0)}=60^{\circ}$ (\color{green}{$\blacksquare$}\color{black}) and $\theta_p^{(0)}=30^{\circ}$  (\color{gray}{$\blacksquare$}\color{black}). The absence of significant differences indicates that $\theta_p^{(0)}$ does not impact the steady state particle motion.\\ 
Finally, we study the motion of neutral ($\theta_c=90^{\circ}$) active particles that approach the interface from the aqueous phase, when the oxygen is more concentrated 
in oil ($E<0$), as displayed in Fig.\ref{Fig:Ludwig_interface-bounce}B. 
%As a further case study, we place particles away from the interface, in the low solute concentration region (water, blue area), moving towards the %interface, as displayed in Fig.\ref{Fig:Ludwig_interface-bounce}B). 
We consider both uniform 
($\mu_+=\mu_-=\mu=0.5$, red) and Janus-like phoretic mobilities ($\mu_+=0.5, \mu_-=0.3$ (black) and $\mu_+=0.3, \mu_-=0.5$ (yellow)).\\ 
%In all cases  particles get trapped next to the interface, acquire a new angle and, in some cases, start moving along the interface. \\
Janus particles tend to reorient such to minimize the 
interfacial overlap of the more repulsive side (larger $\mu$) with the solute-rich liquid.
Thus, the particle with higher front mobility ($\mu_+ > \mu_-$, in black) faces 
the water phase (depicted as a blue area in the inset), attaining a value 
of the orientation angle $\theta_p \approx -90^{\circ}$, 
whereas the opposite occurs when 
$\mu_+ < \mu_-$ ($\theta_p \approx 90^{\circ}$, in yellow).
%solute concentration (oil, yellow area), while particles with softer mobility in the rear %(black particles) face the opposite. Because they align with their director vector, $\hat{m}$, %perpendicular to the interface,  activity cannot move particles along the interface. Moreover, %this force cannot displace  particles away from the interface and hence particles get trapped.
In both cases, since their director vector $\hat{m}$ is othogonal to the interface and 
phoretic forces cannot overcome the capillary trapping, they get stuck. 
On the contrary, for uniform diffusiophoretic mobilities (red particle), interfacial alignement is lacking, and particles displace along the interface.

\section{Conclusions}\label{sec:conclusions}
In this work we have introduced a new model based on Lattice-Boltzmann to study the interaction of active particles with liquid-liquid interfaces. The model facilitates  the study of the full hydrodynamics of the system, on the same footing as  diffusiophoretic and wetting forces suspended particles are subject to. The model allows to switch on and off easily these forces, and to modify the particle properties such as wetting and the diffusiophoretic force, differentiating for this last scenario two parts on the particle with its own activity and mobility. These contributions are formulated locally, and can then be adapted  to particles of arbitrary shape, with a general inhomogeneous  treatment of their surfaces. Moreover, the liquid mixture  can show asymmetric solubility to the chemicals produced by the particles.  \\
We have  tested the interaction of inactive particles trapped at the interface under different wetting angles $\theta_c$ (0º, 90º and 180º) and different particle surface mobility. We have seen that while  wetting dominates over diffusiophoretic forces, when the wetting is neutral (90º), diffusiophoretic properties are important, and inactive particles with homogeneous surface mobility displace from the interface. This interaction is proportional to the surface mobility, and to the different of products between both phases. When the surface has an asymmetric mobility, particles reorient to have its more repulsive face towards the liquid phase with less product, and displace from the phase of high accumulation of product. The reorientation depends on the strength of the mobility. The more repulsive, the fastest reorient. \\
Active particles at the interface with neutral wetting move along the interface. The more asymmetry between product accumulation in both phases and the more repulsive is their surface to products, the fastest particles move. Particles reorient themselves to a specific angle, no matter the angle at which particles are placed. However, if the asymmetry of products between both phases decays, particles find different equilibrium positions depending on the initial angle. This effect is seen for different surface mobilities. If wetting is changed, particles will stay closer to the interface rather if they would not have the activity. If particles with high wetting for the side of initial motion move towards the interface, they will contact the interface, reorient and move along the interface. Depending on the ratio of the surface mobilities, particles can stop, or continue their motion.\\
Overall, the proposed model   has huge capabilities to explain many phenomena occurring at these interfaces, and that sets a new start line where to study these and more complex systems.

\section*{Acknowledgments}
L.P. would like to thank MINECO for the FPI BES-2016-077705 fellowship.
A.S. acknowledges support from the European Research Council under the European
Union Horizon 2020 Framework Programme (No. FP/2014-2020), 
ERC Grant Agreement No. 739964 (COPMAT).
I.P. acknowledges support from Ministerio de Ciencia, Innovaci\'on y Universidades (Grant
No. PGC2018-098373-B-100/FEDER-EU), DURSI (Grant No. 2017 SGR 884),  SNSF (Project No. $200020_204671$), and EU Horizon 2020 Program (Grant FET-OPEN 766972-NANOPHLOW).

\section*{Data Availability Statement}

The data that support the findings of this study are available from the corresponding author upon reasonable request.

\bibliography{spc-interfaces-arxiv} 

\end{document}